\documentstyle[12pt,aas2pp4,flushrt]{article}

\begin{document}

\title{Comparing Galaxies and Ly$\alpha$ Absorbers at Low Redshift}
\author{Suzanne M. Linder}
\affil{The Pennsylvania State University, University Park, PA 16802}
\authoremail{slinder@astro.psu.edu}

\begin{abstract}
A scenario is explored in which Ly$\alpha$ absorbers at low redshift arise 
from lines of sight through extended galaxy disks, including those of 
dwarf and low surface brightness galaxies.  A population of galaxies is
simulated based upon observed distributions of galaxy properties, and 
the gas disks are modeled using pressure and
gravity confinement.  Some parameter values are ruled out by comparing
simulation results with the observed galaxy luminosity function, and 
constraints may be made on the absorbing cross sections of 
galaxies.  Simulation results indicate that it is difficult to match 
absorbers with 
particular galaxies observationally since absorption typically occurs at
high impact parameters ($>200$ kpc) from luminous galaxies.  Low impact
parameter absorption is dominated by low luminosity dwarfs.  A large 
fraction of absorption lines is found to originate from low surface brightness 
galaxies, so that
the absorbing galaxy is likely to be misidentified.  Low redshift Ly$\alpha$
absorber counts can easily be explained by moderately extended galaxy disks
when low surface brightness galaxies are included, and it is easily possible
to find a scenario which is consistent with observed the galaxy luminosity 
function, with low redshift Lyman limit absorber counts, and with standard 
nucleosynthesis predictions of the baryon density, $\Omega_B$.
\end{abstract}


\section{Introduction}

\pagestyle{myheadings}
\markboth{\sc Linder \hfill Comparing Galaxies and Ly$\alpha$ Absorbers~~~}
         {\sc Linder \hfill Comparing Galaxies and Ly$\alpha$ Absorbers~~~}

The numerous absorption lines seen shortward of hydrogen Ly$\alpha$ in 
quasar spectra are a powerful probe of gas in galaxies.  Recent 
observations of the Ly$\alpha$ forest at low redshift (Bahcall et al. 1996) 
have allowed for
studies which compare absorbers with nearby galaxies, but the nature
of the lower column density Ly$\alpha$ absorbers remains controversial.
Here I explore the possible connection of Lyman limit
systems ($N_{HI}=10^{17.2}$ to $10^{20.3}$ cm$^{-2}$) and Ly$\alpha$ forest
absorbers ($N_{HI}<10^{17.2}$ cm$^{-2}$) with galaxies at low redshifts by
using simulations to relate properties of the galaxy population to absorbers.
A complete 
understanding of  galaxy formation and evolution will require knowledge of the 
properties of the low redshift galaxy population, including galaxies which
are more difficult to detect.
Absorbing galaxies should include dwarf and low surface brightness (LSB) galaxies.
These objects are generally found to be rich in gas, so they must make at 
least some 
contribution to Ly$\alpha$ absorption.
Numerous studies have found extended gaseous envelopes around dwarf galaxies, 
so that their absorbing cross sections might not be small as their name 
would imply.

Burbidge et al. (1977) first noted that known 
galaxies must have huge cross sections in order to
explain the Ly$\alpha$ absorbers.  However, galaxy disks
may be quite extended when ionized gas is considered, and the absorption 
capabilities of the more recently
discovered dwarf and LSB galaxies have not been explored.  
Recent studies have had some
success in matching absorption lines with particular galaxies, although 
conclusions vary depending upon the column density threshold.  Studies 
concentrating on high column density, low redshift absorbers
generally find clear 
associations with galaxies (Lanzetta et al. 1995a; Stocke et al. 1995;
Steidel et al. 1997),
while lower column density absorbers tend to be located far from any observed
galaxy.  Often one or more galaxies will be seen at several hundreds of kpc
from the line of sight, so that it is unclear whether any of the galaxies are
responsible for the absorption, although the absorbers are generally found
to follow the large scale galaxy distribution 
(Le Brun, Bergeron, \& Boiss\'e 1996; Morris et al. 1993; Bowen, Blades,
\& Pettini 1996; Rauch, Weymann, \& Morris 1996; Mo \& Morris 1994; Salzer 1992;
van Gorkom et al. 1996).  A few absorbers have been found in cosmic voids
(Stocke et al. 1995; Shull, Stocke, \& Penton 1996), while limited attempts have
been made to detect LSB galaxies near quasar lines of 
sight (van Gorkom et al. 1996).
Several groups have tested for a physical connection between absorbers and
galaxies by looking for an anticorrelation between the equivalent widths of
the absorption lines and the impact parameters between the galaxies and
lines of sight (Lanzetta et al. 1995a; Bowen et al. 1996; Le Brun et al. 1996).
However, it is not possible to know that any particular absorption line
has been matched to the correct galaxy, and some galaxies which are 
unidentified may be causing absorption.  Thus it is also useful to estimate
how many absorbers could arise from lines of sight through galaxies.

Many nearby galaxies have been discovered only recently due
to a severe selection effect against LSB galaxies (Bothun, Impey \& McGaugh
1997).  Disney (1976) made the first quantitative analysis of this selection 
effect after realizing that galaxies obeying the Freeman (1970) law were
within one magnitude of the typical night sky brightness. 
The primary goal of 
recent surveys for LSB galaxies has been simply to obtain a representative
sample of the nearby galaxy population (Schombert et al. 1992; Impey et al. 
1996; Sprayberry et al. 1996; Dalcanton et al. 1997a; O'Neil, Bothun, \& 
Cornell 1997).   Most LSB galaxies 
exhibit a wide range of properties such as sizes and morphologies which are
similar to those of high surface brightness (HSB) galaxies.  The exceptions
are Malin objects, which make up about 10\% of the Schombert et al. (1992)
catalog.
These objects are highly extended and extremely low in surface brightness, and
they are often among the most luminous galaxies known.  While most surveys
for LSB galaxies have been biased toward blue objects, varying colors have
been observed by O'Neil et al. (1997).  McGaugh (1996a)
found that the distribution of central galaxy surface brightnesses is 
approximately flat to 
at least $\mu_0=$ 24 $B$ mag arcsec$^{-2}$.
Evidence has been seen more recently for a falloff faintward of 
24 $B$ mag arcsec$^{-2}$ 
(O'Neil  et al. 1997; 
Zwaan, Briggs, \& Sprayberry 1997; Briggs 1997).
O'Neil et al. (1997) attribute this effect to an inability of extremely
low surface brightness galaxies to form in a cluster environment.

The gaseous extent of galaxies is highly uncertain.  Several studies involving
21 cm mapping of galaxy disks have found 'sharp edges,' where the neutral
hydrogen column density falls off quickly from a few times $10^{19}$ cm$^{-2}$
to an undetectable level.  Such 'edges' have been explained successfully by
models where the ionization level increases rapidly from the optically thick
to optically thin regime (Maloney 1993; Corbelli \& 
Salpeter 1993; Dove \& Shull 1994), thus making a  sudden cutoff in the total
column density of the gas unnecessary.  A detection of ionized gas beyond 
the HI disk in NGC 253 has been made recently (Bland-Hawthorn, Freeman, \&
Quinn 1997).  Ionized outer disks could be quite
extended if they are not severely affected by galaxy interactions.  These 
extended structures may also be consistent with results of cosmological
simulations, where galaxies form within extended sheets and filaments
(Weinberg, Katz, \& Hernquist 1997; Norman et al. 1997).  Many absorbers are
found to arise in the sheet structures which have already 
formed at high redshifts.
Double line of sight observations (Dinshaw et al. 1994, 1995; Fang et al. 1996)
have shown that
absorbers are hundreds of kpc in size.  Furthermore, these observations were
found to be consistent with flattened absorbers having organized motion,
such as rotating disks (Charlton, Churchill, \& Linder 1995).
Disks in reality are likely to be warped and broken, but when considering
the average behavior of a large sample of galaxies, some
systematic falloff in total column density with radius can be expected.
Only field galaxies are considered here, without the effects of clustering
and interactions.

The number density of galaxies at low redshift is also uncertain due to a
possible abundance of LSB galaxies.  An upper limit can be found by 
supposing that all absorbers are caused by galaxy disks.  The average number
of absorbers per unit redshift along a line of sight was found by Bahcall
et al. (1996) to be $(dN/dz)_0=24.3\pm 6.6$, complete to an equivalent width in
Ly$\alpha$ of 0.24 \AA. The neutral hydrogen column density 
distribution is observed
to be approximately a power law at high redshifts 
(but see Petitjean et al. (1993) and
Charlton, Salpeter, \& Linder (1994)) with a slope of -1.5.
While an accurate distribution of neutral hydrogen column densities at low
redshift has not been published yet, it is assumed to look similar to
those at higher redshifts.  A significant 
change in slope is generally not observed between 
redshifts 4 and 2.7 for the column density range of interest here 
(Lu et al. 1996; Hu et al. 1995; Kirkman \& Tytler 1997; Kim et al. 1997).

Section 2 of this paper describes the method, 
including defining and simulating galaxy
populations and modeling the galaxy disks as absorbers.  Results are discussed
and compared with local luminosity functions, the predicted baryon density,
and absorber/galaxy observations in section 3.
Future papers will discuss the effects of clustering and the possibility of
distinguishing between this scenario and and others, such as those with 
significant nongalactic absorbers.
The value of $H_0=100 $ km s$^{-1}$ Mpc$^{-1}$ is assumed.

\section{Method}

Samples of galaxies were simulated from a galaxy population as defined in 
section 2.1,  where the method for choosing galaxy luminosities and surface
brightnesses is explained.  In each simulation, all of the galaxies were 
given random coordinates within a cube of space with a side length of 50 Mpc.
Each galaxy disk was modeled using pressure and gravity 
confinement, as described in section 2.2.  Lines of sight, going
straight through the cube, were then chosen randomly.  When a
line of sight intersected a galaxy disk, the neutral column density was found
by integrating the neutral gas density along the line of sight.

\subsection{Defining a Galaxy Population}

A galaxy population is defined here using simple parameter distributions
which are consistent with observations.  Galaxy luminosities are determined
by the choice of scale lengths and central surface brightnesses.
Relations are then assumed between these observed parameters and the properties
of the gas disks.

The surface brightness distribution
is assumed to be flat with an exponential falloff at the high surface 
brightness (HSB) end, as in McGaugh (1996a).
The LSB end is known to extend to at least 24 $B$ mag arcsec$^{-2}$, while
a few objects have been observed with $\mu_0 >$27 $B$ mag arcsec$^{-2}$.
Surface brightness profiles for the individual galaxy disks are assumed to be
exponential, where the distribution of observed scale lengths is found to
be a power law, $\phi (h_B)\propto h_B^n$, where $n\sim -2$ (van der Kruit 1987;
Hudson \& Lynden-Bell 1991; de Jong 1995).
No correlation is assumed between scale length and central
surface brightness, except that large, HSB objects are known not to exist
(Sprayberry et al. 1995).  Bothun, Impey \& Malin (1991) have
shown that the above distributions require a slope of $-1.5$ in the faint end
of the galaxy luminosity function.  

The bright end of the luminosity function
is generally well fit by a Shechter function (Marzke, Huchra, \& Geller 1994).
Undiscovered Malin objects cause some incompleteness (McGaugh 1996a), 
although they
are thought to be rare (Dalcanton, Spergel, \& Summers 1997b; Hoffman, Silk,
\& Wyse 1992).  Additional faint galaxies have been added in evolutionary
models which avoid 
strong evolution in the faint blue galaxy population at higher redshifts
(Gronwall \& Koo 1995; McGaugh 1994; Driver et al. 1995).  
Sprayberry et al. (1997) found that a combined Shechter
bright end and power law faint end ($M_B>-16$) provide the best fit for
the observed luminosity 
function, including LSB galaxies, where the faint end slope was 
found to be quite steep ($\sim -2$).
Observed luminosity functions (Driver \& Phillips 1996; Zucca et al. 1996; 
Marzke et al. 1994) are generally found to be consistent with such hybrid
shapes, including a Shechter function with $\alpha =-1$ and a
power law with slope between $-1.4$ and $-1.8$ for $M_B>-16$, as discussed
in Sprayberry et al. (1997). Here the disk scale length 
distribution,  for a given $\mu_0$, obeys a power law if $M_B > M_{Bsw}$.
For more luminous galaxies, a scale length is chosen, for a 
given central surface
brightness, so that the Marzke et al. (1994) Shechter function is obeyed.

LSB galaxies obey the Tully-Fisher relation (Zwaan et al. 1995),
and simulations by Navarro (1997) have hinted at a physical explanation for
this relation.
Here the Tully-Fisher formula from Jacoby et al. (1992) is used to determine a 
rotation velocity for each galaxy, which is needed to model the dark matter 
halo.  Evidence is given by de Blok,
McGaugh, \& van der Hulst (1996) that LSB galaxies are low in 
central column density compared to HSB objects.
A central column density value for HSB galaxies has been estimated by 
Bowen, Blades,
\& Pettini (1995).  A relationship is assumed here which is consistent with
the Bowen et al. (1995) estimate and which requires a factor of five change
in surface brightness for a factor of two in central column density 
(McGaugh 1996b) so that the central column density becomes
$N_0=10^{-0.17\mu_0+25.9}$ cm$^{-2}$.

\subsection{Modeling Galaxy Disk Absorbers}

The total (neutral plus ionized) column density is assumed to fall off 
exponentially in each inner disk and as a power law farther out:
\begin{equation}
%\end{array}
N_{tot}(r)=\left\{ \begin{array}{l@{\quad\mbox{for}\quad}l}
N_0\exp (-r/h_{21}) & r<r_{sw}\\ N_0\exp (-r_{sw}/h_{21})(r/r_{sw})^{-p} &
r>r_{sw}
\end{array}
\right. 
\end{equation}
where $r$ is the distance in the plane of the disk, $r_{sw}$ is the value of 
$r$ at which the switch from exponential to power law occurs, and $h_{21}$
is an exponential scale length of the inner neutral hydrogen disk which 
could be determined using 21 cm mapping.
Evidence for a power law falloff in outer galaxy disks has been seen by 
Hoffman et al. (1993), and such behavior is required for galaxies to produce
a power law type neutral column density distribution.  A neutral column
density distribution which goes as $N_H^{-1.5}$ corresponds to $p=4$.  The
value of $r_{sw}$ is quite uncertain, and here it is assumed to be at one or two
times the radius at which the gas becomes highly ionized ($r_{cr}$).  The 
ratio of the neutral hydrogen scale length to $B$ scale length ($h_{21}/h_{B}$)
is also uncertain.
An initial estimate of 3.4 is used here based upon the conjectures of 
Salpeter (1995), and a smaller value (1.7) is also tested.

The disks are modeled using pressure and gravity confinement (Charlton, 
Salpeter, \& Hogan 1993; Charlton et al. 1994).  The vertical
structure is calculated in each disk, so that the internal pressure and density
are related by $P(z,r)=2n_{tot}(z,r)kT$, where the assumed temperature  is
$T=20000$ K, $z$ is the height above the disk plane, and $n_{tot}$ is the 
density of neutral plus ionized hydrogen.  
The inner disks are confined by a spherical dark halo with
gravitational acceleration $g=V_{rot}^2z/r^2$ where
$V_{rot}$ is the rotation velocity.  The boundary condition 
$P(z=w/2)=P_{ext}$ defines the 
absorber width $w$.  The outer disks are confined by an
external pressure $P_{ext}$ which may be due to infalling material or to 
a local
or diffuse intergalactic medium. The high redshift value of $P_{ext}/k=10$
cm$^{-3}$ K \newpage
(Charlton et al. 1994) is used here, although more recently 
evidence has emerged supporting
a similar value at low redshift (Wang \& Ye 1997).
Hydrostatic equilibrium is assumed, so that
\begin{equation}
\frac{dP}{dz} = -\frac{m_HP}{2kT}\frac{V_{rot}^2z}{r^2}
\end{equation}
where $m_H$ is the mass of a hydrogen atom. The solution is
\begin{equation}
P(z)=P(0)\exp (z^2/h_z^2),
\end{equation}
where
\begin{equation}
P(0)=P_{ext}\exp (w^2/4h_z^2),
\end{equation}
is the pressure at $z=0$, and
\begin{equation}
h_z=\frac{2r}{V_{rot}}\left(\frac{kT}{m_H}\right)^{1/2}
\end{equation}
is the vertical scale height.

A model for the vertical ionization structure is used which is similar
to that in Maloney (1993).   The gas is assumed to be highly ionized above 
height $z_i$
where
\begin{equation}
\int_{z_i}^{w/2}\alpha_{rec}n_{tot}^2(z)dz = \phi_{i,ex},
\end{equation}
and $\phi_{i,ex}$ is the ionizing flux, which has been measured at
low redshift by Kulkarni \& Fall (1993) using the proximity effect.
The inner layer of gas below $z_i$ is shielded and remains neutral.
This sandwich structure occurs out to a fairly constant column density of 
about $3\times 10^{19}$ cm$^{-2}$ for each galaxy.  For $r>r_{cr}$ ,
where the column
density is below this critical value, ionization equilibrium is assumed, where
$\alpha_{rec}n_{tot}^2=\zeta n_H$
for neutral hydrogen density $n_H$, $\zeta$ is an average
ionization rate, and $\alpha_{rec}=4.07\times 10^{13}$ cm$^3$ s$^{-1}(T/10,000
\mbox{K})^{-0.75}$ is the recombination coefficient.

For lines of sight intersecting the inner disk,
the observed column density is
dominated by the neutral layer of gas, so that
\begin{equation}
N_{HI}=\frac{P(0)}{2kT}\int_{\mbox{los}}\exp (-z^2/h_z^2)dz
\end{equation}
where the integral is performed along the line of sight (los) for $|z|<z_i$.
For highly ionized gas ($r>r_{cr}$) the neutral column density becomes
\begin{equation}
N_{HI}
=\frac{\alpha_{rec}}{\zeta} \frac{P(0)}{2kT}\int_{\mbox{los}}
\exp (-2z^2/h_z^2)dz.
\end{equation}

\section{Results and Discussion}
The first set of simulations each consisted of 20,000 lines of sight through 
a box containing 20,000
galaxies.  The following standard parameters were used for each simulation, 
with the exceptions listed in Table 1.  
\begin{itemize}
\item The slope of the neutral column density 
distribution was $\epsilon =1.5$ (1.7 for run 1).
\item The slope of the $B$ scale length distribution was
$n=-2$ ($-1.35$ for run 2).
\item The cutoff in the surface brightness distribution was
at $\mu_0=$ 24 $B$ mag 
arcsec$^{-2}$ (25 for run 3; 27 for run 9) .
\item The ionizing flux  $\phi_{i,ex}=4.6\times 10^3
$ cm$^{-2}$s$^{-1}$ ($2.8\times 10^4$ cm$^{-2}$s$^{-1}$ for run 4).
\item The ratio of the 21 cm scale length to B scale length was 
$h_{21}/h_B=3.4$ (1.7 for run 5).
\item The external pressure was 
$P_{ext}/k=10$ cm$^{-3}$ K (1 cm$^{-3}$ K for run 6).  
\item The switch from exponential to power law falloff in total column density
occurred at $r_{sw}=r_{cr}$ ($r_{sw}=2r_{cr}$ for run 7).
\item The luminosity function was changed from a Shechter form to a power law
at $M_{Bsw}=-16$ ($-15$ for run 8).
\end{itemize}

\subsection{Galaxy Counts and the Cosmological Density Parameter}

The number density of galaxies for each simulation, 
listed in Table 1, is that which
is required to explain the observed absorber counts of $(dN/dz)_0=24.3\pm 6.6$
(Bahcall et al. 1996) for $N_{HI}>10^{14.3}$ cm$^{-2}$.  For
galaxies with relatively large absorption cross sections (run 1) fewer galaxies
are required, while the opposite effect is seen in runs 4, 5, 6, and 7.  Since 
the central column density of the disk is assumed to 
vary rather weakly for a large 
change in surface brightness, LSB galaxies have bigger absorption cross 
sections at a given $M_B$ compared to HSB galaxies, so that fewer galaxies are
required if there are more LSB objects (runs 3 and 9).  Absorption cross 
sections are generally found to increase with $M_B$, however, so that runs (2
and 8) where the shape of the luminosity function is changed to favor more 
luminous objects require fewer galaxies.  Note that not all dwarf galaxies are
required to be rich in gas as in these models, as the number density of 
galaxies is relatively insensitive to the faint end slope of the luminosity
function.  Some dwarfs may have formed recently or by varying mechanisms 
(Hunsberger, Charlton, \& Zaritsky 1996).

\begin{figure}[tb]
\plotone{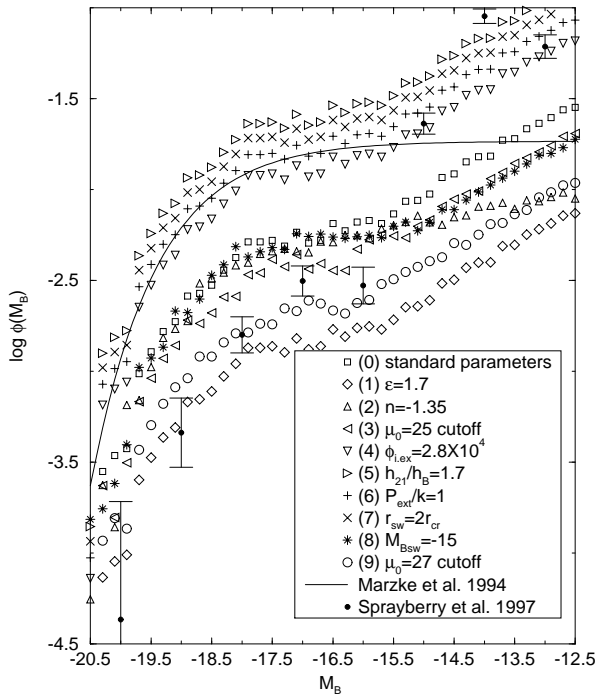}
\figcaption[fig1.eps]{Simulated galaxy luminosity functions, which are 
normalized to produce observed absorber counts, are plotted along with 
observed LSB (binned points) and HSB (solid line) luminosity functions.
Note that most of the runs shown here require fewer galaxies than are known
to exist in order to explain the observed value of $(dN/dz)_0$.
\label{fig1}}
\end{figure}

The simulated luminosity functions (LFs), 
which are normalized to produce absorber
counts, are shown in Figure~\ref{fig1}.  The Shechter function found by Marzke et al.
(1994) and the binned LF from Sprayberry et al. (1997) are also shown.
The Marzke et al. (1994) function generally includes galaxies with $\mu_0<$
22 $B$ mag arcsec$^{-2}$, while the Sprayberry et al. (1997) points are 
restricted to $\mu_0>$ 22 $B$ mag arcsec$^{-2}$, so that
the sum of the two can be considered to be a fairly complete LF.  Note that
only models with relatively small galaxy absorption cross sections can provide
a reasonable fit to the observed LF.  In other words, models where galaxies
have larger cross sections will overproduce $dN/dz$ if they contain as many
galaxies as are known to exist according to observed LFs.  
Therefore, by comparing with
observed LFs, it is possible to rule out models with overly large galaxy
cross sections, or put an upper limit on the characteristic size of
absorbing galaxies.  For example, a galaxy with $M_B^*=-18.9$ would have a
radius of around 100-300 kpc in runs 4 and 5 down to a column density of 
$2\times 10^{14}$ cm$^{-2}$, where some variations occur due to variations
in central surface brightness.  Smaller galaxy disks cannot be
ruled out by this method, but then significant nongalactic absorption must
also occur. Many parameter assumptions are also uncertain.
For example, several galaxies with $\mu_0\sim$ 27 $B$ mag arcsec$^{-2}$
are known to exist,
so there must be at least some tail at the LSB end of the surface brightness
distribution.  In reality, some adjustments or variations in all of the 
parameters are expected.  It can be concluded, however, that it is easily 
possible to 
obtain a reasonable fit to the observed luminosity function, and thus explain
absorbers using galaxies, within the range of parameter uncertainties.

A value for the cosmological density parameter $\Omega_{DM}$, 
due to the given number density
of galaxy halos required to match the observed $(dN/dz)_0$, is also shown 
in Table 1.  It is assumed that a flat rotation
curve continues to three times the radius of the pressure-gravity confinement
transition in each galaxy.  Zaritsky et al. (1997) find that 
luminous spiral galaxy halos continue out to at least 200 kpc, which is 
consistent with the radius assumed for a typical $L^*$ galaxy in run 4 or
5.
The value of $\Omega_{DM}$ is generally found to be larger for models with a
greater number density of required galaxies,  although varying the shape of
the faint end of the LF has little effect (runs 2 and 8).  Simulations which 
produce enough galaxies compared to the
observed LF tend to have high values of $\Omega_{DM}$.
However, the value of $\Omega_{DM}$ is  most sensitive to the 
shape of the bright end of the LF and the properties of the halos, which
need to be modeled more accurately.

The value of $\Omega_B$ due to hydrogen in galaxies, which is also shown in
Table 1, is more closely related to absorber counts.  Here $\Omega_B$ is
shown due to galaxies assuming each one cuts off at a neutral column density 
of $2\times 10^{14}$ cm$^{-2}$,
although varying the cutoff column density
in each run has little effect.  Assuming each galaxy has associated gas out to
a column density of $10^{12}$ cm$^{-2}$ increases $\Omega_B$ by less than
$\sim 1\%$, while assuming that the gas cuts off at the pressure-gravity 
confinement transition decreases $\Omega_B$ by $<10\%$.  Previous estimates
of $\Omega_B$ due to observed galaxies at low redshift have been much 
smaller than those from primordial nucleosynthesis
predictions (Persic \& Salucci 1992), but comparing with 
the value from Walker et al. (1991), $\Omega_Bh_{100}^2=0.013\pm 0.003$, most
of the runs give baryon densities which are somewhat high.  
One exception is run 1,
which also requires an unreasonably small number of galaxies as discussed 
above.  Runs which give particularly high $\Omega_B$ include 4, 6 and 7.
In run 4, a large amount of ionized gas is required, and in run 7, the gas 
is too highly concentrated in the centers of galaxies.  In run 6, the disks 
must be confined by gravity out to larger radii, and a more rapid falloff in 
neutral column density occurs in the gravity dominated regime (Charlton et al.
1994), so again the gas is too centrally concentrated. 
It can be argued from the 
results of run 7 that galaxies must have surrounding gas which falls off 
slowly with radius (such as a power law rather than an exponential) in order 
for galaxies to explain most of the Ly$\alpha$ absorbers. 
The only remaining way to allow for a reasonable fit to the LF is to decrease
the 21 cm scale lengths as in run 5, although the value of $\Omega_B$ remains
slightly high.

\begin{figure}[tb]
\plotone{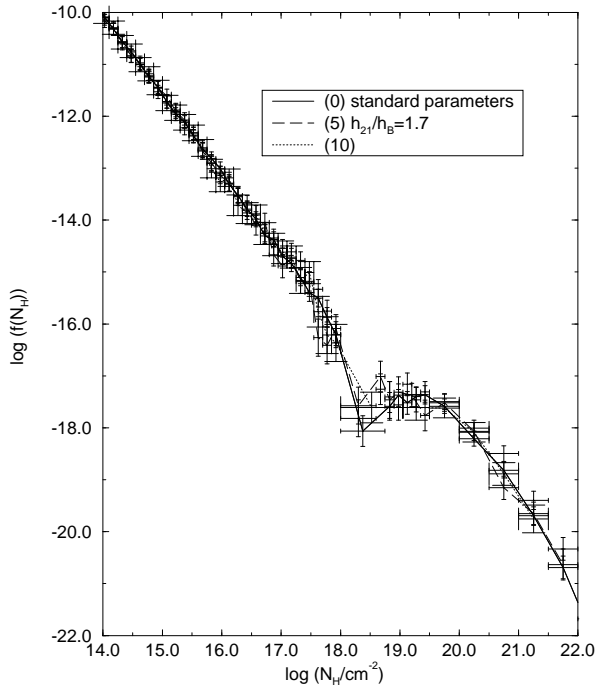}
\figcaption[fig2.eps]{Simulated neutral column density distributions for
low redshift Ly$\alpha$ absorption.  Their shapes are determined primarily 
by the assumed slope, and therefore, and observed neutral column density 
distribution will be useful primarily for constraining this slope. \label{fig2}}
\end{figure}

It is possible that the amount of gas in galaxy centers has been overestimated
by assuming exponential total column density profiles, since the observed 
central column densities tend to be about an order of magnitude smaller than
the extrapolated fits to exponential profiles (Cayatte et al. 1994; de Blok et
al. 1996).  Most of this unobserved gas mass is likely to exist in the form of
molecular gas or stars (McGaugh \& de Blok 1997) so that $\Omega_B$ is not 
severely overestimated.  A somewhat high value of $\Omega_B$ obtained from 
the simple models used here cannot be taken too seriously for a given set of
parameters, and small 
adjustments of multiple parameters can reduce the value of $\Omega_B$.  For 
example, adding more LSB galaxies would be reasonable, 
as some tail must exist at
the LSB end of the surface brightness distribution, but the shape of 
this tail is highly uncertain.  Also, decreasing $r_{sw}/r_{cr}$ might make 
the models more realistic.

The number of Lyman limit absorbers is shown for each simulation in Table 1, again
assuming that the total number of absorbers from Bahcall et al. (1996) is 
produced by galaxies.  The number of Lyman limit absorbers has been measured
at redshifts $\ge 0.36$, and the lowest redshift values available are within
the range of $(dN/dz)_{LL}\sim 0.2$ to 1.3 (Lanzetta, Wolfe, \& Turnshek 
1995b; Storrie-Lombardi et al. 1994; Stengler-Larrea et al. 1995).  
Storrie-Lombardi et al. (1994) find 
a power law evolution in the number of Lyman limit systems which decreases 
at lower redshifts, while the Lanzetta et al. (1995b) and Stengler-Larrea et
al. (1995) measurements are less sharply
decreasing with redshift.  Assuming that the evolution is flat or decreasing
to $z\sim 0$, the values of $(dN/dz)_{0,LL}$ are slightly high for most runs.
Runs 4, 6, and 7, which also had unreasonably high $\Omega_B$, have excessive 
numbers of Lyman limit absorbers.  Thus the same model parameters seem to be
favored by comparing with Lyman limit absorber counts as with $\Omega_B$, as
discussed above.

Since there are likely to be small variations
in all of the model parameters, no attempt is made to find a single 
best fitting set of parameters.
Further discussions concentrate on the relatively reasonable run 10, which 
uses the standard parameters except for a surface brightness cutoff at 25 $B$
mag arcsec$^{-2}$, a scale length ratio of $h_{21}/h_B=1.7$, and a change to
a power law falloff in column density at $r_{sw}=0.8r_{cr}$.  In run 10,
20,000 galaxies and 30,000 lines of sight are simulated, and a number density of
2.0 galaxies Mpc$^{-3}$ is required to explain absorber counts, giving 
$\Omega_{DM}=0.37$, $\Omega_B=0.017$, and $(dN/dz)_{0,LL}=1.56$.  Note that
the Lyman limit systems are now produced in regions with power law falloffs in
column density.  The number
of Lyman limit systems would be smaller if the column density profiles were 
modeled more realistically, so that the Lyman limit regime is produced in 
a region with column density falloff which is faster than the power law, yet
slower than the exponential.

\subsection{Implications for Ly$\alpha$ Absorbers}

\begin{figure}[tb]
\plotone{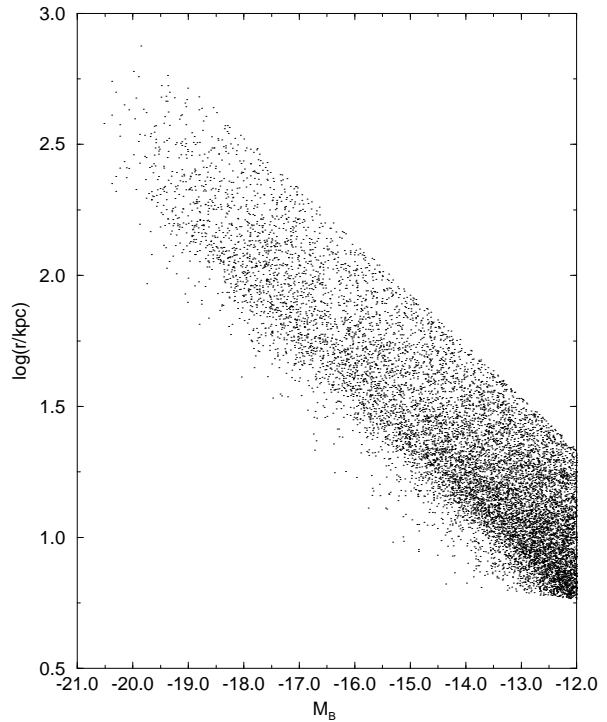}
\figcaption[fig3.eps]{The absorbing ($>2\times 10^{14}$ cm$^{-2}$) radii of 
galaxies (simulated in run 10) increase with galaxy luminosity, although
variation occurs at a given $M_B$ which depends upon surface brightness. 
\label{fig3}}
\end{figure}

\begin{figure}[thb]
\plotone{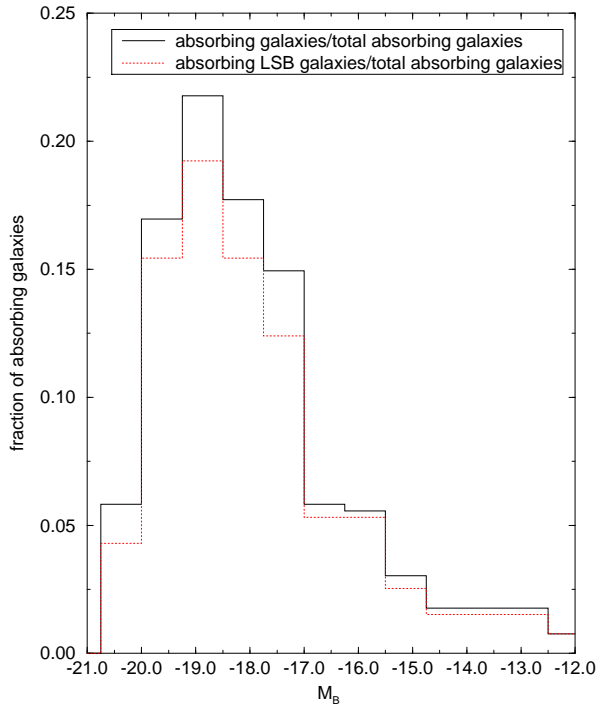}
\figcaption[fig4.eps]{The fraction of absorbing ($>10^{15}$ cm$^{-2}$) galaxies
at impact parameters $< 300$ kpc is shown as a function
of $M_B$ for run 10. LSB galaxies, with $\mu_0>$ 22 $B$ mag arcsec$^{-2}$, 
may cause most of the 
Ly$\alpha$ absorption.  Low luminosity dwarfs make a modest contribution 
to absorbing galaxies. \label{fig4}}
\end{figure}

Simulated neutral column density distributions are shown in Figure~\ref{fig2} 
(for larger simulations consisting of 
30,000 galaxies and 30,000 lines of sight in runs 0 and 5).  Little
variation is seen between different simulations since the normalization is fixed and the slope
is not expected to vary much from $-1.5$.  A measurement of the observed slope
at low redshift
will be interesting, as the value of $\Omega_B$ is sensitive to small 
changes in $\epsilon$, although constraints
on the other model parameters are unlikely to be possible using an observed 
neutral column density 
distribution.  A gap is seen in each simulation at $\sim 10^{19}$ cm$^{-2}$ due to
the sudden increase in ionization, although it may be smoothed out in
reality due to clumpiness in the gas or variations in the amount of 
Ionizing radiation.

\begin{figure}[htb]
\plotone{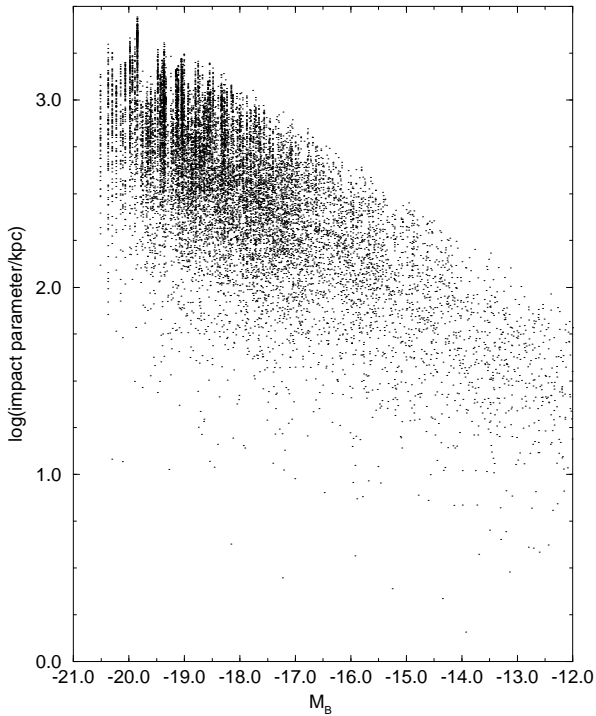}
\figcaption[fig5.eps]{Impact parameters between lines of sight and absorbing
($>10^{12}$ cm$^{-2}$) galaxies in run 10 versus $M_B$.  Luminous galaxies
can absorb out to larger impact parameters as is consistent with Figure 3.
Note that most objects
at low impact parameters are low in luminosity. 
%Vertical stripes appear because
%individual galaxies (especially luminous ones) can cause absorption along 
%multiple lines of sight. 
\label{fig5}}
\end{figure}

Absorption cross sections are found to increase with galaxy luminosity, as
shown in Figure~\ref{fig3}.  Thus most absorbers will arise from luminous
galaxies, although less luminous objects may produce some absorption since
they are more numerous.  It was also shown by Bowen et al. (1996) that 
absorption cross sections must depend upon galaxy luminosity.  
They argued that it is unlikely that low luminosity galaxies
produce absorption due to this dependence, but a modest fraction of absorption
at low impact parameters is produced by galaxies which are low in luminosity, 
as shown in Figures~\ref{fig4} and~\ref{fig5}.
Variations in
surface brightness have less effect upon absorption cross section, but at
a given $M_B$ a lower surface brightness galaxy will be larger since the
column density is assumed to vary weakly with surface brightness.  Thus the
upper points in Figure~\ref{fig3} show a sharp edge due to the unrealistic sharp
cutoff assumed in the surface brightness distribution, 
while the lower (HSB) points 
fall off more smoothly.  

A surprisingly large fraction of absorbers ($>10^{15}$
cm$^{-2}$) is shown
to arise in LSB galaxies, as is seen in Figure~\ref{fig4}.  
While this result is quite 
sensitive to the shape of the surface brightness distribution and to the 
relationship between column density and surface brightness, luminous HSB 
galaxies may make a relatively unimportant contribution to Ly$\alpha$ 
absorption.  While LSB galaxies are thought to have relatively low central
column densities, a strong correlation is not seen between column densities
and surface brightnesses in de Blok et al. (1996).  Making this relationship
weaker results in a larger contribution to absorption from LSB galaxies.
In Figure~\ref{fig5}, the 
absorbing ($>10^{12}$ cm$^{-2}$) galaxy impact parameters are shown versus
$M_B$.  More luminous galaxies can absorb at larger impact parameters, as is
consistent with Figure~\ref{fig3}.  It can also be seen that low luminosity galaxies
make an important contribution to absorption if they are close to lines of
sight and that a large fraction of absorbers arise in somewhat luminous LSB
galaxies which allow for absorption at the largest impact parameters.  Radio
surveys for absorbing galaxies such as van Gorkom et al. (1996)
find that absorbers include dwarf galaxies at relatively small impact 
parameters and luminous galaxies which are farther from lines of sight.

\begin{figure}[tbh]
\plotone{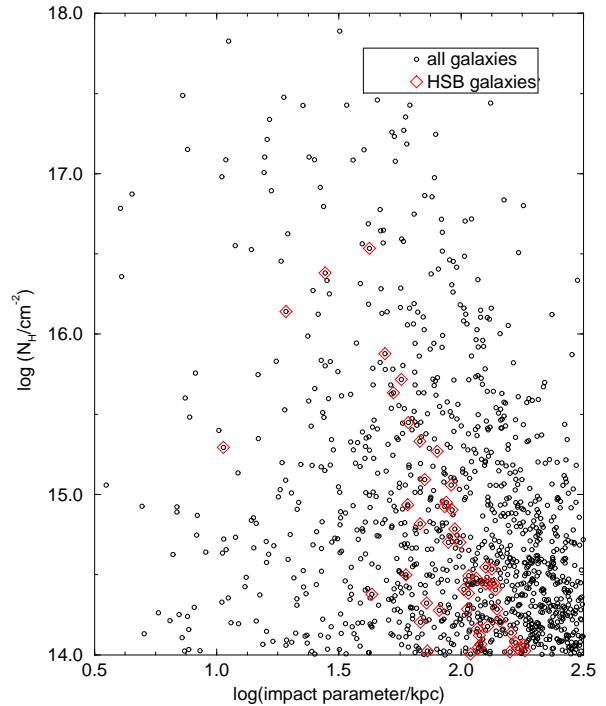}
\figcaption[fig6.eps]{Impact parameters between absorbing galaxies and lines
of sight in run 10 are plotted versus neutral column density for all galaxies
(circles) and for HSB ($\mu_0<$ 22 $B$ mag arcsec$^{-2}$) galaxies (diamonds).
\label{fig6}}
\end{figure}

\begin{figure}[tb]
\plottwo{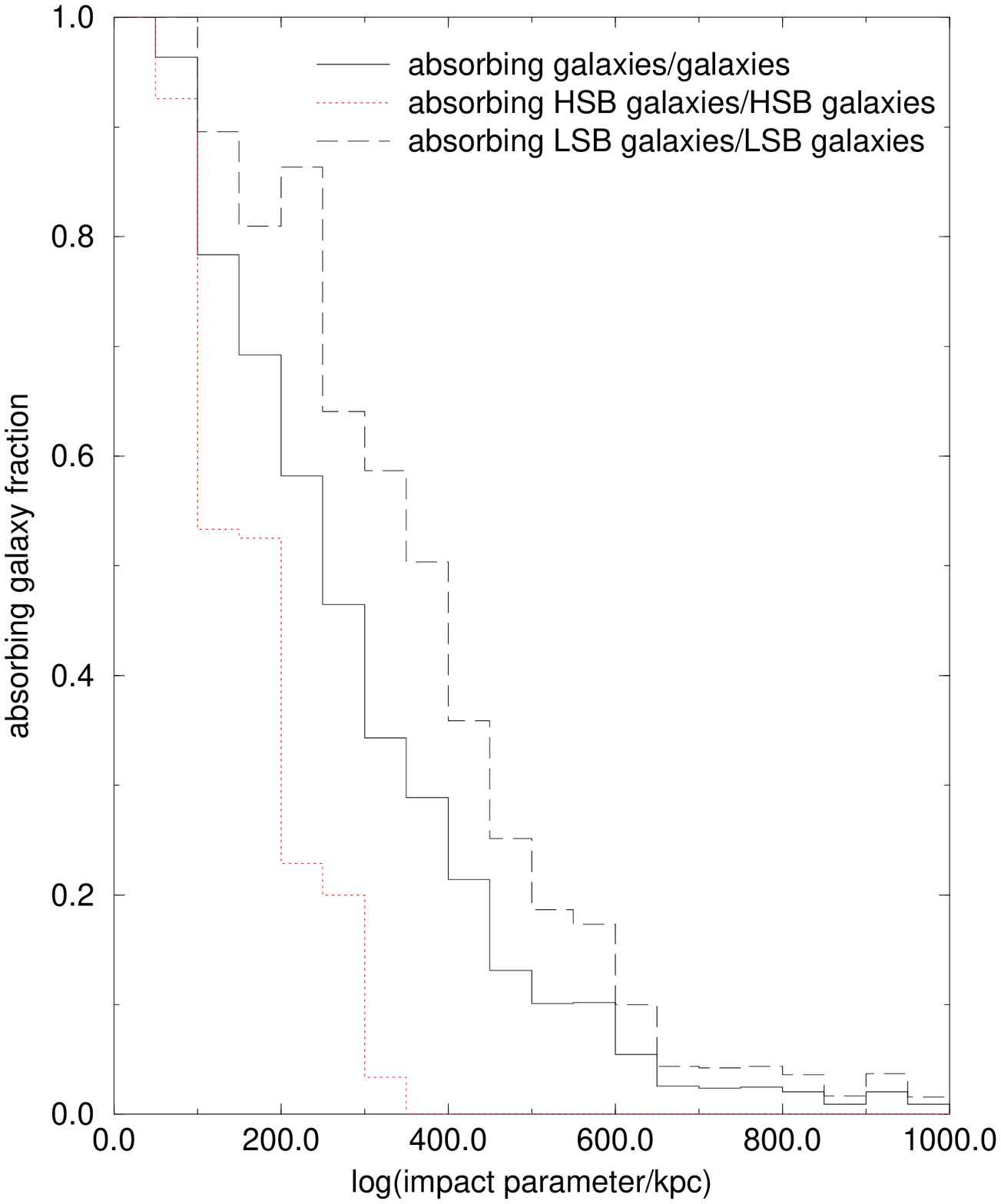}{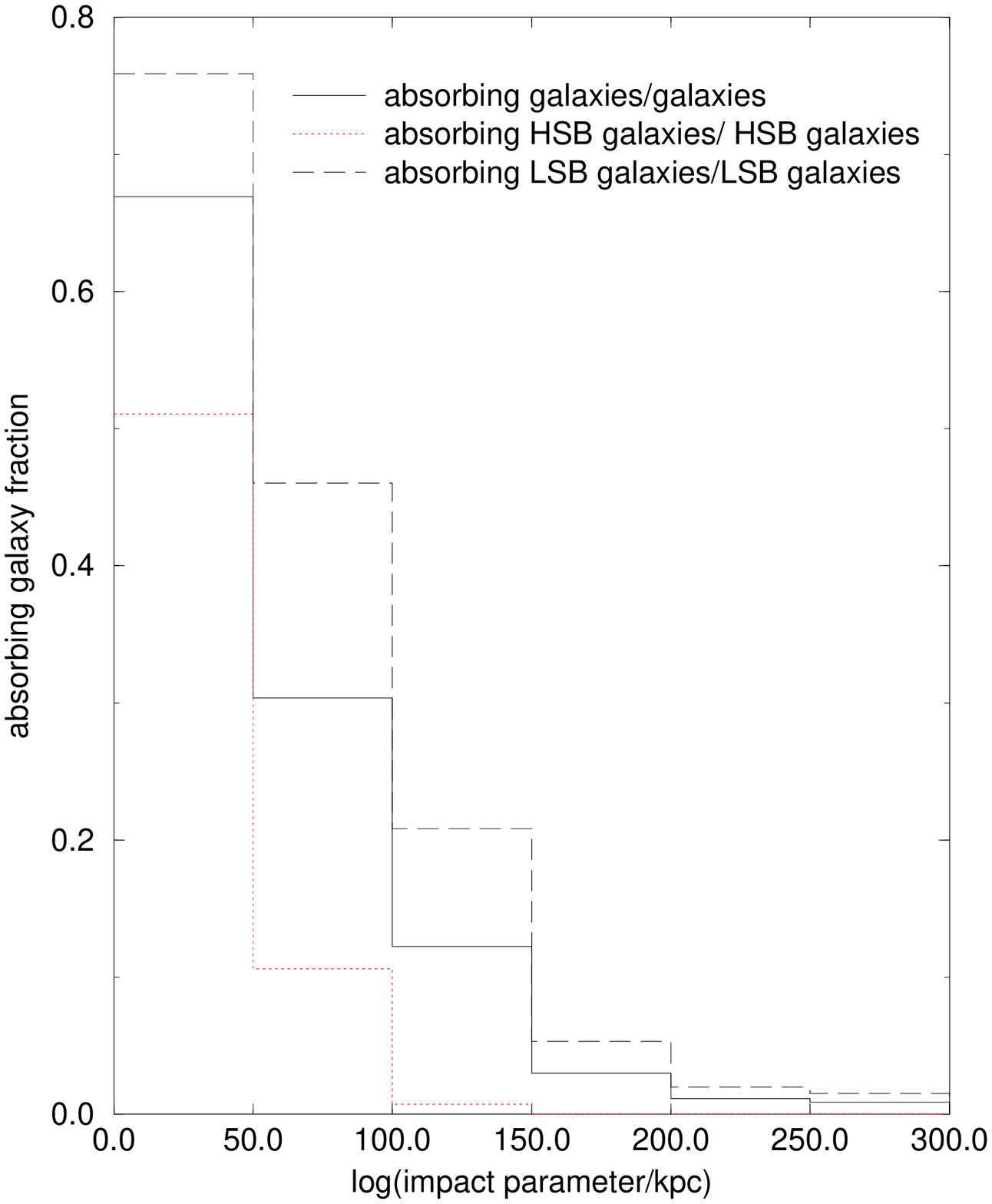}
\figcaption[fig7.eps]{The covering factors, or fractions of galaxies causing 
absorption in a given impact parameter range are shown for run 10.  
The fraction of
absorbing LSB ($\mu_0>$ 22 $B$ mag arcsec$^{-2}$, dashed line) and HSB 
($\mu_0<$ 22 $B$ mag arcsec$^{-2}$, dotted line)
galaxies in a given range of impact parameter which cause absorption are also
shown separately. In Fig. 7a (7b)
the plotted galaxies have luminosities 
$>L^*$ ($0.1L^*$) and cause absorption $>10
^{14.3}$ cm$^{-2}$ ($10^{15}$ cm$^{-2}$).  LSB galaxies are
more likely to cause absorption as they have larger cross sections at a given
luminosity. \label{fig7}}
\end{figure}

A plot of the galaxy impact parameters versus neutral column density is shown
in Figure~\ref{fig6}.   While each galaxy is assumed to have a systematic falloff in  
column density with radius, a large amount of scatter is seen in the
plot mostly due to variations in absorption cross sections of galaxies.
Variations in disk inclinations contribute less to the scatter, so that
any scenario in which absorption cross
sections vary would produce a similar plot.  However, when realistic
selection effects are introduced, limiting the galaxy sample to a more
narrow range of properties, less scatter is seen.  A strong statistical 
anticorrelation is found for all (963) points simulated with $N_{HI}>10^{14.3}$
using a Kendall rank correlation test.
However, when random samples of 30 points are chosen, a $>2\sigma$ 
anticorrelation is found in only 122 out of 200 samples.  Furthermore, those
samples which are strongly anticorrelated typically contain one high column
density ($>10^{20}$ cm$^{-2}$) system.
Lanzetta et al. (1995a) observed an anticorrelation between
equivalent width and corresponding galaxy impact parameter.
Bowen et al. (1996) saw
no obvious anticorrelation, 
which was used to argue that absorbers are more weakly
associated with galaxies.  Le Brun et al. (1996) observed luminous
galaxies close to lines of sight, to a lower equivalent width limit.  They
found a weak anticorrelation between equivalent width and impact parameter.
All of these results appear consistent with the scenario described here,
and many more observations will be required to distinguish between absorber
models.

The fraction of galaxies which cause absorption is shown versus galaxy impact
parameter in Figures~\ref{fig7}a ($L>L^*$, $N_{HI}>10^{14.3}$ cm$^{-2}$) and 
~\ref{fig7}b 
($L>0.1L^*$, $N_{HI}>10^{15}$ cm$^{-2}$).  Bowen et al. (1996) find that 
their observed
galaxies typically cause absorption out to 300 kpc, while Lanzetta et al.
(1995a) find a smaller typical absorption cross section of about 160 kpc. This
value is quite consistent with cross sections found for HSB, $M_B^*$ galaxies.
Lanzetta et al. (1995a) find
that 5/5 galaxies with impact parameters $<70$ kpc cause absorption, where
these galaxies are slightly less luminous than the ones plotted in Figure~\ref
{fig7}a.
Le Brun et al. (1996) find that all
of 4 galaxies cause absorption within $\sim 90$ kpc where their selection 
effects are similar to those in Figure~\ref{fig7}a.
Bowen et al. (1996) find that 40 percent of galaxies cause absorption at 
impact parameters between 100 and 300 kpc, although their sample is not quite
complete to $0.1L^*$ as in Figure~\ref{fig7}b.  While selection effects which depend
upon redshift may explain some of the discrepancies, it is also likely that
dwarf and LSB absorbers are being misidentified.  In the simulation results 
here, the correct absorbing galaxy is known, while observers are likely to
match an absorption line with the most easily visible galaxy.

The clustering properties of the galaxies would need to be included in the
models in order to test whether galaxy misidentifications
could help to explain the results from Bowen et al. (1996).  LSB galaxies 
may be weakly clustered around galaxies which are more easily seen, so that
the absorption cross sections of some luminous galaxies may be overestimated.
Many galaxies which are more difficult to detect may have formed within the
same extended sheets which contain luminous, HSB galaxies as seen in 
cosmological simulations.  It is also 
possible that a more clumpy distribution of gas around galaxies, such as
that produced by tidal interactions (Morris \& van den Bergh 1994), would be
more realistic, although the covering factor of gas around the galaxies would
need to remain high out to larger distances in order to be consistent with 
observations.  Both Morris et
al. (1993) and Le Brun et al. (1996) found that all luminous galaxies within 0.5
to 1 Mpc cause absorption with a lower equivalent width limit.  It would be
difficult to allow for individual galaxies having such large cross sections
without overproducing absorber counts.

\section{Conclusions}

Ly$\alpha$ absorber counts at low redshift can easily be explained by lines
of sight through moderately extended galaxy disks when low surface brightness
galaxies are included. A scenario in which absorbers are explained by galaxies
can easily be made consistent with observed luminosity functions, low redshift
Lyman limit absorber counts, and predictions of $\Omega_B$ from standard
nucleosynthesis.
In successful models the absorption cross sections 
must increase
with galaxy luminosity, where an $L^*$ galaxy typically has an absorbing radius
of $\sim 200$ kpc and a low luminosity dwarf has a radius $\sim 10$ kpc.  
Larger absorption cross sections are ruled out by 
observed luminosity functions, assuming that most galaxies are potential
absorbers.  The baryon density, as predicted by primordial 
nucleosynthesis, is overproduced if the gas is too highly ionized or too
concentrated in the centers of galaxies, while low surface brightness 
galaxies are less centrally concentrated, so that they can have larger 
absorption cross sections without containing excessive baryons.

Low surface brightness galaxies may make a large and important
contribution to Ly$\alpha$ absorption.  In fact, luminous, high surface
brightness galaxies may make a relatively unimportant contribution.
Luminous galaxies, especially those 
which are low in surface brightness, are responsible
for most Ly$\alpha$ absorption lines, where the lower column densities are 
often due to luminous galaxies at large impact parameters.  Low luminosity
dwarfs are much more numerous, so that their contribution to absorption 
becomes more important when they are at smaller impact parameters from quasar 
lines of sight.  Therefore, it may be generally difficult to observationally 
match a particular absorption line to the correct galaxy.  

While a fairly small number of Ly$\alpha$ absorption lines have been 
matched to specific galaxies to date, these observations generally appear
consistent with the scenario simulated here, given that it is impossible
to know that any particular absorption line has been matched to the correct
galaxy.  A substantial fraction of Ly$\alpha$ 
absorption is likely to arise from lines of sight through low surface 
brightness galaxies which are clustered around more easily visible galaxies.
These LSB galaxies are likely to be misidentified, and improved simulations
which consider the clustering properties of these galaxies will be needed to
quantify the effects of possible misidentifications.
A much larger number of observations will be required to test possible models
(Sarajedini, Green, \& Jannuzi 1996), and greater sensitivity to low surface
brightness galaxies near quasar lines of sight will be required, even in 
places where luminous, high surface brightness galaxies have already been
detected.
Refinements to the models here which allow
for reasonable clustering properties will allow for further constraints on
the arrangement of neutral and ionized gas relative to galaxies in the 
nearby Universe.

\acknowledgments

I am grateful to J. Charlton, G. Bothun, C. Churchill, R. Ciardullo, 
S. McGaugh, J. Miralda-Escud\'e, and D. Schneider
for valuable discussions and to D. Sprayberry for supplying
data points.  This work was supported by a fellowship from the NASA Graduate
Student Researchers Program.

\clearpage

\begin{deluxetable}{cccccc}
\tablecaption{Simulation Results}
\tablehead{
\colhead{run \#} & \colhead{parameters varied} & \colhead{galaxies (Mpc)$
^{-3}$} & \colhead{$\Omega_{DM}$} & \colhead{$\Omega_B$} & \colhead{$(dN/dz)
_{0,LL}$}}
\startdata
0 & all standard & 0.62 & 0.21 & 0.021 & 1.59 \nl
1 & $\epsilon =1.7$ & 0.16 & 0.06 & 0.006 & 0.415 \nl
2 & $n= -1.35$ & 0.28 & 0.19 & 0.020 & 1.39 \nl
3 & $\mu_0=25$ cutoff & 0.52 & 0.16 & 0.018 & 1.31 \nl
4 & $\phi_{i,ex}=2.8\times 10^{4}$ & 1.4 & 0.50 & 0.049 & 2.25 \nl
5 & $h_{21}/h_B=1.7$ & 2.8 & 0.53 & 0.023 & 1.41 \nl
6 & $P_{ext}/k=1$ & 1.9 & 1.00 & 0.063 & 4.77 \nl
7 & $r_{sw}=2r_{cr}$ & 2.3 & 0.44 & 0.077 & 4.04 \nl
8 & $M_{Bsw}=-15$ & 0.44 & 0.21 & 0.021 & 1.34 \nl
9 & $\mu_0 = 27$ cutoff & 0.39 & 0.11 & 0.014 & 1.22 \nl
\enddata
\tablecomments{Columns are: (3) the number density of galaxies
required to explain absorber counts, (4) the cosmological density parameter
due to dark matter halos (out to three times the gravity confinement radius 
in each galaxy), (5) the cosmological density parameter due to hydrogen (out 
to a column density of $10^{14.3}$ cm$^{-2}$), and (6) the number of Lyman 
limit systems per unit redshift.}
\end{deluxetable}
\end{document}